\documentclass[twocolumn,showpacs,preprintnumbers,amsmath,epsfig,apsrev,aps,prb]{revtex4}

\usepackage{graphicx}
\usepackage{graphics}
\usepackage{dcolumn}
\usepackage{bm}

\begin{document}
\preprint{}

\title{Inelastic neutron scattering and frequency domain magnetic resonance studies of \emph{S}=4 and \emph{S}=12 Mn$_6$ single-molecule magnets}

\author{O.~Pieper$^{1,2}$}
\email[]{oliver.pieper@helmholtz-berlin.de}
\author{T.~Guidi$^{3}$}
\email[]{tatiana.guidi@stfc.ac.uk}
\author{S.~Carretta$^{4,5}$}
\author{J.~van Slageren$^{6,7}$}
\author{F.~El Hallak$^{6}$}
\author{B.~Lake$^{1,2}$}
\author{P.~Santini$^{4,5}$}
\author{G.~Amoretti$^{4,5}$}
\author{H.~Mutka$^{8}$}
\author{M.~Koza$^{8}$}
\author{M.~Russina$^{1}$}
\author{A.~Schnegg$^{1}$}
\author{C.J.~Milios$^{9}$}
\author{E.K.~Brechin$^{9}$}
\author{A.~Juli\`{a}$^{10}$}
\author{J.~Tejada$^{10}$}

\affiliation{$^{1}$Helmholtz-Zentrum Berlin f\"{u}r Materialien und Energie (HZB), Glienicker Stra\ss e 100, 14109 Berlin, Germany\\
$^{2}$Institut f\"{u}r Festk\"{o}rperphysik, Technische Universit\"{a}t Berlin, Hardenbergstra\ss e 36, 10623 Berlin, Germany\\
$^{3}$ISIS Facility, Rutherford Appleton Laboratory, Chilton, Didcot, Oxon OX11 0QX, UK\\
$^{4}$Dipartimento di Fisica, Universit\`{a} and Unit\`{a} CNISM di
Parma, I-43100 Parma, Italy\\
$^{5}$National Research Center on nanoStructures and bioSystems at
Surfaces (S3), CNR-INFM, 41100 Modena, Italy\\
$^{6}$1. Physikalisches Institut, Universit\"{a}t Stuttgart, D-70550 Stuttgart, Germany\\
$^{7}$School of Chemistry, University of Nottingham, Nottingham NG7 2RD, United Kingdom\\
$^{8}$Institute Laue-Langevin, B.P. 156, F-38042 Grenoble Cedex, France\\
$^{9}$University of Edinburgh, West Mains Road, Edinburgh, EH9 3JJ, United Kingdom\\
$^{10}$Departament de F\'isica Fonamental, Facultat de F\'isica,
Universitat de Barcelona, Avinguda Diagonal 647, Planta 4, Edifici
nou, 08028 Barcelona, Spain}

\begin{abstract}
We investigate the magnetic properties of three Mn$_6$ single
molecule magnets by means of inelastic neutron scattering and
frequency domain magnetic resonance spectroscopy. The experimental
data reveal that small structural distortions of the molecular
geometry produce a significant effect on the energy level diagram
and therefore on the magnetic properties of the molecule. We show
that the giant spin model completely fails to describe the spin
level structure of the ground spin multiplets. We analyze theoretically the spin Hamiltonian for the low spin Mn$_6$ molecule ($S=4$) and we show that the excited $S$ multiplets play a key role in determining the effective energy barrier for the magnetization reversal, in analogy to what was previously found for the two high spin Mn6 ($S=12$) molecules [S. Carretta \textit{et al.}, Phys. Rev. Lett. \textbf{100}, 157203 (2008)].
\end{abstract}


\date{\today}

\maketitle

\section{Introduction}
\begin{figure}[htb!]
\includegraphics[width=8cm,angle=0]{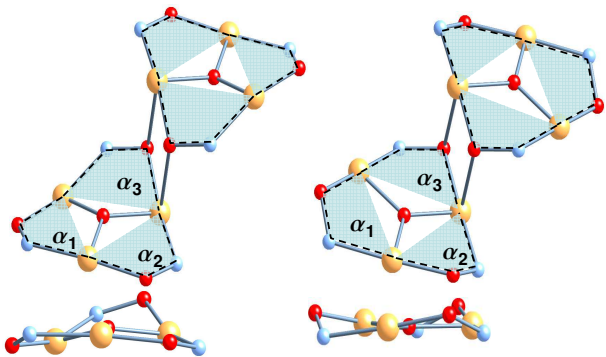}
\caption{(color online) Core of molecules (\textbf{1}) (left) and (\textbf{3}) (right) showing at the bottom the difference in torsion angles ($\alpha_1$, $\alpha_2$ and $\alpha_3$). Color scheme: Mn, orange, O, red,  N, blue. H and C ions are omitted for
clarity.} \label{fig:Mn6_S4_S12}
\end{figure}

Single Molecule Magnets (SMMs) have been the subject of intense
research activity since the first and mostly studied one,
Mn$_{12}$-ac, was reported \cite{Sessoli93}. These metal-organic
clusters are usually characterized by a large spin ground state S
and an easy-axis anisotropy which determines the Zero-Field
Splitting (ZFS) of the $S$ state sublevels. The resulting magnetic
bistability makes them interesting for magnetic storage applications
due to their potential to shrink the magnetic bit down to the size
of one single molecule. Until recently and despite the common
efforts of chemists and physicists to find suitable systems that
could retain the magnetization for a long time at non cryogenic
temperatures, Mn$_{12}$-ac was the system showing the `highest'
blocking temperature (3.5 K) and anisotropy barrier (74.4
K)\cite{Chackov06}. The relaxation time in the classical regime
follows the Arrhenius law: $\tau = \tau_0 \exp(U/k_BT)$ (Ref.
\onlinecite{Villain94}). According to this, there are two key points
that have to be considered for the realization of an ideal SMM.
First of all, the anisotropy barrier, given to a first approximation
by $U\sim |D|S^{^2}$ ($D$ is the axial anisotropy parameter), has to
be sufficiently high. This is to prevent the reversal of the
magnetization via a classical thermally activated multistep Orbach
process mediated by spin-phonon interactions. This can be achieved
by the simultaneous increase of $D$ and $S$, two variables that are
intrinsically linked together \cite{Waldmann07}. Secondly, the
pre-exponential factor $\tau_0$ in the Arrhenius law has to be
large. This factor is dominated by the time necessary to climb the
upper states in the energy level diagram, and is proportional to
$D^{-3}$ (Ref. \onlinecite{Villain94}). In addition to the classical
relaxation mechanism, the quantum tunneling of the magnetization
(QTM) that characterizes the spin dynamics of SMMs, has to be taken
into consideration and minimized for magnetic data storage application, since it provides a
shortcut for the relaxation of the magnetization.\\
Therefore, to engineer SMMs able to retain the magnetization for
long time it is crucial to control all the different mechanisms that
provide a relaxation path for the system. Recently we succeeded in
the synthesis of a new class of Mn$^{3+}$-based clusters that
contributed in raising the anisotropy barrier and has served as a
good model system to study the factors involved in the relaxation
mechanism \cite{Carretta08, Carretta09poly}. \\This class consists of
hexanuclear Mn$^{3+}$ clusters (from now on Mn$_6$) which, despite
the generally similar nuclear structure, display a rich variety of
spin ground states and anisotropy energy barriers
\cite{Milios07Ueff53, Milios07Ueff86, Milios07switch,
InglisDalton09, MiliosDalton08,Milios07magstruc}. The six Mn$^{3+}$
ions are arranged in two triangles, with dominant ferromagnetic (FM)
exchange interaction between the two triangles and FM or
antiferromagnetic (AFM) interactions within the two triangles. It
has been found that the nature of the intra-triangle exchange
interaction can be switched from AFM to FM by substituting the
organic ligands bridging the Mn$^{3+}$ ions, leading to a change of the
ground state from a low spin ($S=4$) to a high spin ($S=12$)
\cite{Milios07Ueff53}. Furthermore, deliberately targeted structural
distortions have been successfully used to tune the values of
the exchange interactions \cite{Milios07Ueff86}. The isotropic exchange
interactions, and consequently the overall anisotropy barrier
\cite{Carretta08}, is thus found to be very sensitive to the
structural details. This has been also demonstrated using an
alternative method for distorting the molecule, that is by applying
external hydrostatic pressure and correlating the structural changes
with the magnetic behavior \cite{Prescimone08}. It is therefore
quite important to determine the exchange interactions for different
structures to deduce magneto-structural correlations. This
information can be then used to engineer new clusters with
selectively modified molecular structures that match
the optimized conditions for the desired magnetic properties. \\
We have investigated three members of the family of Mn$_6$ clusters,
with chemical formulas
[Mn$_6$O$_2$(sao)$_6$(O$_2$CMe)$_2$(EtOH)$_4$]$\cdot$4EtOH
(\textbf{1}),
[Mn$_6$O$_2$(Et-sao)$_6$(O$_2$CPh)$_2$(EtOH)$_4$(H$_2$O)$_2$]$\cdot$2EtOH
(\textbf{2}) and
[Mn$_6$O$_2$(Et-sao)$_6$(O$_2$CPh(Me)$_2$)$_2$(EtOH)$_6$]
(\textbf{3}) \cite{Milios04,Milios07Ueff53,Milios07Ueff86}. All
molecules display very similar structures consisting of six
Mn$^{3+}$ ions ($s = 2$) arranged in two staggered triangular units
(see Fig. \ref{fig:Mn6_S4_S12}) related by an inversion centre.

The only major structural difference between the three clusters
resides is the steric effect of the organic ligands used in proximity to the transition
metal ions. However, despite having very similar structures, the
three molecules have very different magnetic properties. The
coupling between the magnetic ions occurs via superexchange pathways
involving oxygen and nitrogen ions and is found to be extremely
sensitive to intramolecular bond angles and distances. The
particular arrangement of the magnetic ions provides exchange
couplings lying in the cross-over region between AFM and FM. For
this reason, even small structural distortions have tremendous
impact on the magnetic properties of the system. For example, while
the coupling between the two triangles is ferromagnetic for all
molecules, the intra-triangular coupling changes from
antiferromagnetic in (\textbf{1}) to ferromagnetic in (\textbf{2})
and (\textbf{3}) due to a 'twisting' of the oximate linkage. This results
in a 'switching' of the total spin ground state from $S=4$
to $S=12$. Systematic synthesis and studies of various members of the
Mn$_6$ family have revealed that the nature of the coupling is
extremely sensitive to the intra-triangular Mn-O-N-Mn torsion angles
\cite{MiliosDalton08,InglisDalton09} (see Fig.
\ref{fig:Mn6_S4_S12}). There is a critical value for the torsion
angle of $30.85\pm0.45^{\circ}$, above which the pairwise exchange
interaction switches from antiferromagnetic to ferromagnetic, while
a further enhancement of the angle increases the strength of the FM
interaction. This effect has been interpreted in terms of the
particular arrangement of the manganese d$_{z^2}$ orbitals with
respect to the p-orbitals of the nitrogen and oxygen ions. A large
(small) Mn-O-N-Mn torsion angle results in a small (large) overlap
between the magnetic orbitals giving rise to ferromagnetic
(antiferromagnetic or weak ferromagnetic) superexchange interactions
\cite{Cremades09}.\\
Molecules (\textbf{2}) and (\textbf{3}) have the same spin ground
state $S=12$ but very different effective energy barriers
($U_{\text{eff}}\approx 53$ K for (\textbf{2}) and
$U_{\text{eff}}\approx 86.4$ K for (\textbf{3})). This difference
was found to be closely related
to the exchange interactions \cite{Carretta08}.\\
In order to understand this rich variety of behaviors, we performed
a detailed spectroscopic characterization of the three molecules
using inelastic neutron scattering (INS) and frequency domain
magnetic resonance (FDMR).  FDMR is only sensitive to transitions
with a predominantly intra-multiplet character, according to the
selection rules $\Delta S=0, \Delta M_S=\pm1$. In contrast, in INS
both inter- and intra-multiplet transitions can be observed ($\Delta
S=0,\pm1, \Delta M_S=0,\pm1$). Thus, the combination of the two
techniques allows assignment of all observed excitations
\cite{Sieber05,Piligkos05}.

The determination of the model spin Hamiltonian parameters enabled
us to estimate theoretically the effective energy barrier for the low spin molecule
(\textbf{1}). Similarly to what we previously reported for the two high spin molecules (\textbf{2}) and (\textbf{3}), the results on (\textbf{1}) show how the presence of low-lying excited spin multiplets plays a crucial role in determining the relaxation of the magnetization. 

In conventional systems, the effects of S-mixing can be effectively
modeled by the inclusion of fourth order zero-field splitting
parameters in the giant spin Hamiltonian \cite{Liviotti02}. Here we will show that
this Hamiltonian is completely inadequate for the description of the
spin state energy level structure.

\section{Experimental Methods}
Non-deuterated polycrystalline samples were synthesized according to
published methods \cite{Milios07Ueff53, Milios07Ueff86}.

FDMR spectra were recorded on a previously described quasi-optical
spectrometer,\cite{vanslageren03} which employs backward wave
oscillators as monochromatic coherent radiation sources and a Golay
cell as detector. Sample (\textbf{1}) proved to deteriorate rapidly
upon pressing and over time. Therefore, the FDMR measurements on
(\textbf{1}) were performed on loose microcrystalline material (348
mg) held between two quartz plates. In this unconventional
measurement, the detector signal was recorded as function of
frequency at different temperatures. Extreme care had to be taken to
prevent the slightest positional changes of sample and equipment,
which changes the standing wave pattern in the beam, precluding
normalization. The normalized transmission was calculated by
dividing the signal intensity at a given temperature by that at the
highest temperature (70K). Sample (\textbf{2}) and (\textbf{3})
deteriorate to a lesser extent and FDMR spectra were recorded on
pressed powder pellets made by pressing ca. 250 mg of the unground
sample, with ca. 50 mg n-eicosane (to improve pellet quality) into a
pellet. All spectra were simulated using previously described
software.\cite{kirchner07}

INS experiments were performed using the multi disc-chopper
time-of-fight spectrometers V3/NEAT at the Helmholtz-Zentrum Berlin
f\"{u}r Materialien und Energie (HZB, Berlin, Germany) and IN5 and
IN6 at the Institute Laue Langevin (Grenoble, France). The samples
were inserted into hollow cylindric shaped Aluminum containers and
mounted inside a standard orange cryostat to achieve a base
temperature of 2 K. A vanadium standard was used for the detector
normalization and empty can measurements were used for the
background subtraction.

\section{Theoretical Modeling and Experimental Results}
The experimental data have been modeled using both the giant spin
Hamiltonian (GSH), which considers the ZFS of the ground state
multiplet only, and the microscopic spin Hamiltonian, which treats
isotropic exchange and single-ion ZFS at the same level. Including
only ZFS terms, the giant spin Hamiltonian for a spin state $S$ reads:
\begin{eqnarray}
H_S = D_S\hat{S}_z^2 + E_S(\hat{S}_x^2-\hat{S}_y^2)+B_4^0\hat{O}_4^0,
    \label{eq:GSH}
\end{eqnarray}
where $D_S$ and $E_S$ are second order axial and transverse anisotropy, respectively, and $B_4^0$ is the fourth order axial anisotropy, with $\hat{O}_4^0$ the corresponding Stevens operator.
The microscopic spin Hamiltonian includes an isotropic exchange term for each pairwise interaction and single ion ZFS terms for each ion:
\begin{eqnarray}
H&=&\sum_{i < j}J_{i j}{\bf s}(i)\cdot {\bf s}(j) + \sum_{i} d_is_z^2(i) + \sum_{i}\biggl( 35 c_i s_z^4(i)\nonumber\\
 &&+ c_i (25 -30 s(s+1)) s_z^2(i)\biggr) ,
    \label{eq:H_micro}
\end{eqnarray}
where ${\bf s}(i)$ are spin operators of the $i^{\text{th}}$ Mn ion. The
first term is the isotropic exchange interaction, while the second
and third terms are the second and fourth order axial single-ion
zero-field splitting, respectively (the $z$ axis is assumed
perpendicular to the plane of the triangle).\\
The spin Hamiltonians have been numerically diagonalized by
exploiting the conservation of the $z$-component of the molecular
total spin and the exchange and anisotropy parameters have been
varied to obtain a best fit of the experimental data.

\subsection{Mn$_6$ (\textbf{1}) ($S=4$) $U_{\text{eff}}\approx$ 28 K}
\begin{figure}[htb!]
\includegraphics[width=8cm,angle=0]{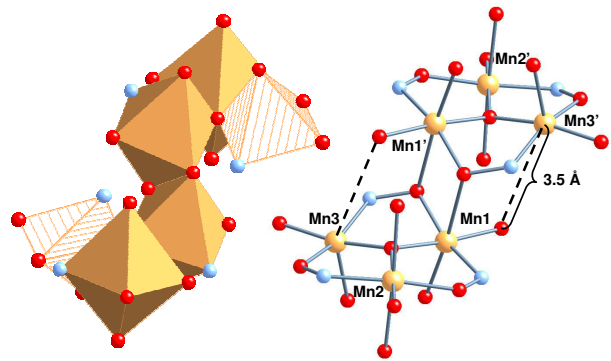}
\caption{(color online) Structure of the Mn$_{6}$ (\textbf{1})
molecular core. The Mn$^{3+}$ ions are located at the vertices of
two oxo-centered triangles. Ions Mn1, Mn2, Mn1' and Mn2' are in
octahedral geometry and ions Mn3 and Mn3' in square pyramidal
geometry, as highlighted in filled and striped orange (left figure).
Color scheme: Mn, orange, O, red,  N, blue. H and C ions are omitted
for clarity.} \label{fig:Mn6_S4_poly}
\end{figure}

Sample (\textbf{1}) was the first reported member of the Mn$_6$
family\cite{Milios04}. The building block of the molecule is the
[Mn$^{3+}_{3}$ O] triangular unit where Mn$_2$ pairs, bridged by the
NO oxime, form a -Mn-O-N-Mn- moiety (Fig. \ref{fig:Mn6_S4_poly}).

\begin{figure}[htb!]
\includegraphics[width=8cm,angle=0]{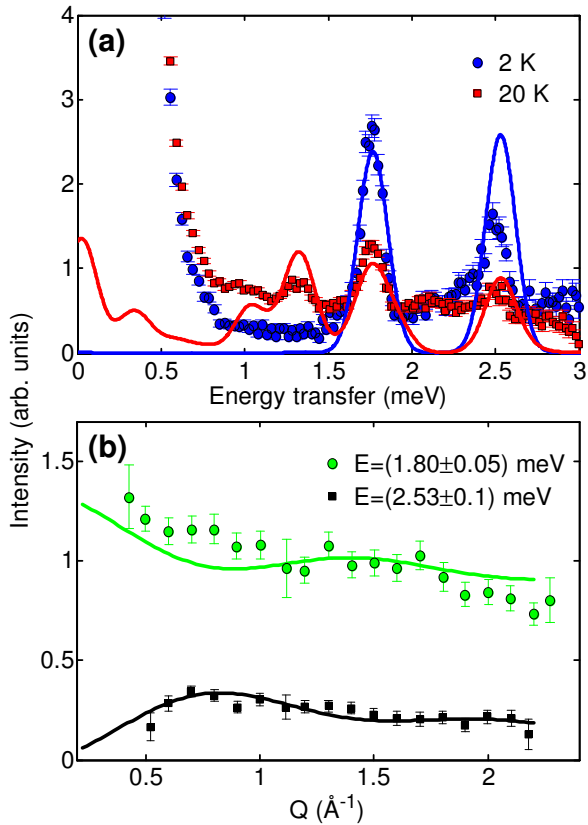}
\caption{(Color online) (a) INS spectra of (\textbf{1}) with an
incident wavelength of $\lambda=4.6$ \AA\ (NEAT) for $T=2$ K (blue
circles) and $T=20$ K (red squares). The continuous lines represent
the spectra calculated assuming a dimer model for the spin
Hamiltonian (Eq. 3). (b) $Q$-dependence of first intra- (green
circles) and inter-multiplet (black squares) transitions measured on
IN6 for $\lambda=4.1 $ \AA\, and $T$=2 K. Continuous lines represent
the calculated $Q$-dependence using the dimer spin Hamiltonian Eq. (\ref{eq:H_dimer})
(assuming a dimer distance of $R=5.17$ \AA, which corresponds to the distance between
the centre of the two triangles).} \label{fig:Mn6_S4}
\end{figure}

The Mn-O-N-Mn torsion angles within each triangle are
10.7$^{\circ}$, 16.48$^{\circ}$ and 22.8$^{\circ}$, giving rise to a
dominant antiferromagnetic exchange coupling \cite{MiliosDalton08}.
The two triangular units are coupled ferromagnetically, resulting in
a total spin ground state of $S=4$. Four out of the six metal ions
(Mn1, Mn2, Mn1', Mn2') are six-coordinate and in distorted
octahedral geometry (MnO$_5$N), with the Jahn-Teller axis almost
perpendicular to the plane of the triangle, while the two remaining
ions (Mn3, Mn3') are five-coordinate and in square pyramidal
geometry (see Fig. \ref{fig:Mn6_S4_poly}). The effective energy
barrier was determined from AC susceptibility measurements to be
$U_{\text{eff}}=28$ K, with $\tau_0=3.6\times10^{-8}$ s (Ref.
\onlinecite{Milios04}). From the effective energy barrier an
estimate of $D\approx -0.15$ meV was derived.\\\indent We performed
INS and FDMR measurements to characterize the ground multiplet and
to identify the position of the lowest-lying excited states from
which we determine the effective exchange interaction and the
zero-field splitting parameters.
\begin{figure}[htb!]
\includegraphics[width=8cm,angle=0]{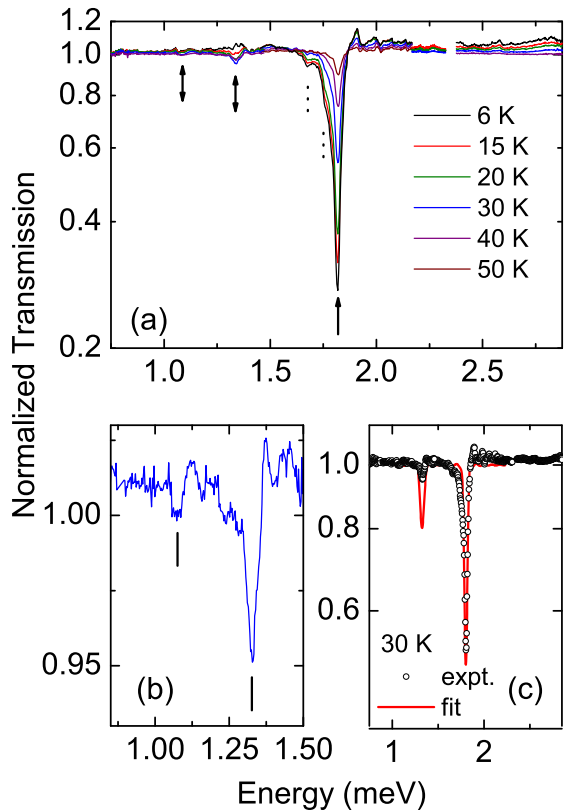}
\caption{(Color online) (a) FDMR spectra of unpressed
polycrystalline powder of (\textbf{1}) recorded at various
temperatures. The intensity of the higher-frequency resonance line
decreases with temperature, while that of the lower-frequency lines
increases up to 30 K, beyond which it decreases again. Dotted lines
indicate resonance lines due to impurities. (b) Expanded view of the
low frequency part of the 30 K spectrum. (c) Experimental and fitted
spectrum at $T$ = 30 K using the GSH approximation. Note the logarithmic scale
in (a) and (c).}
\label{fig:FDMR_Mn6_S4}
\end{figure}
Figure \ref{fig:FDMR_Mn6_S4} shows the FDMR spectra recorded on 350
mg unpressed powder of (\textbf{1}). The most pronounced feature is
the resonance line at 1.803(7) meV, while much weaker features can
be observed at 1.328(1) meV and 1.07(1) meV. The intensity of the
higher-frequency line is strongest at lowest temperature, proving
that the corresponding transition originates from the ground spin
multiplet. The lower-frequency lines have maximum intensity at
around 30 K. No further features were observed between 0.5 and 3
meV. The intense resonance line shows two shoulders to lower
energies, which are much stronger in pressed powder samples and also
increase with the age of the sample. This behavior is mirrored by
the development of a pronounced asymmetric lineshape in INS studies
on older samples. We attribute these shoulders to microcrystalline
particles that have suffered loss of lattice solvent, which leads to
small conformational changes and this alters the ZFS and exchange
parameters. We discard the possibility of isomers with different
orientations of the Jahn-Teller distortion axes, as observed for
Mn$_{12}$\cite{Aubin01}, because we see no signature of different
isomers in the AC susceptibility. We also discount the possibility
of closely spaced transitions, as observed in the Fe$_{13}$ cluster
\cite{vanslageren06}, because the intratriangle exchange
interactions are not equal.\\\indent The higher frequency resonance
line is attributed to the transition from the
$\left|S=4,M_S=\pm4\right>$ to $\left|S=4,M_S\pm3\right>$ states.
INS measurements have found to be necessary to unambiguously
identify the origin of the lower frequency transitions (see below). Assuming
that these transitions are transitions within the ground multiplet,
a fit of the giant spin Hamiltonian ZFS parameters (Eq.
\ref{eq:GSH}) to the observed resonance line energies yields
$D_{S=4}=-2.12\pm0.03$ cm$^{-1}$ ($-0.263 \pm0.004$ meV) and
$B_4^0=+(1.5\pm0.5)\times10^{-4}$ cm$^{-1}$ ($1.24\pm0.06
\times10^{-5}$ meV). This ground state $D_S$-value is much larger
than reported spectroscopically determined $D_S$-parameters for
other manganese SMMs, e.g. $D_{S=10} = -0.457$ cm$^{-1}$ for
Mn$_{12}$Ac \cite{Mirebeau99}, $D_{S=\frac{17}{2}} = -0.247$
cm$^{-1}$ for Mn$_{9}$ \cite{Piligkos05}, or $D_{S=6} = -1.16$
cm$^{-1}$ for Mn$_{3}$Zn$_{2}$ \cite{Feng08}. The main reason for
this large $D$-value is the fact that the projection coefficients
for the single ion ZFS onto the cluster ZFS are larger for spin
states with lower $S$ (Ref. \onlinecite{Bencini90}).
The determined $D_{S=4}=-2.12$ cm$^{-1}$ value for (\textbf{1}) is
in excellent agreement with that found from DFT calculations
($D=-2.15$ cm$^{-1}$) \cite{Ruiz08}. The expected energy barrier
toward relaxation of the magnetization calculated from the found 
spin Hamiltonian parameters is  $U_{\text{theor}}=|D|S^2 = 48.8$ K,
which is much larger than the experimentally found
$U_{\text{eff}}\approx$ 28 K, indicating that more complex
relaxation dynamics characterize this system, in analogy to what has
been found for the Mn$_6$ $S=12$ compounds \cite{Carretta08}. The
linewidth of the 1.33 meV line is slightly larger than that of the
1.80 meV line (48 $\mu$eV versus 41 $\mu$eV), which can indicate the
presence of more than one excitation. The simulated spectrum agrees
very well for the higher-frequency resonance line (note that the
intensity is not rescaled), while the lower-frequency line is much
weaker in the experiment than from the fit. This can be tentatively
attributed to the presence of low-lying excited states as observed
previously for Mn$_{12}$Ac \cite{vanslageren09}.  To determine the
energy of excited spin states and identify the origin of the low
frequency resonances we resorted to INS, the technique of choice to
directly access inter-multiplet excitations.
\\\indent The INS experiments were performed on $\approx 4$ g of
non-deuterated polycrystalline powder of (\textbf{1}), which was
synthesized as described in Ref.\onlinecite{Milios04}. For our
measurements we used incident neutron wavelengths ranging from 3.0
\AA\ to 5.92 \AA\
with energy resolution between 50 $\mu$eV and 360 $\mu$eV.\\
Figure \ref{fig:Mn6_S4} (a) shows the INS spectra for an incident
wavelength of 4.6 \AA\ collected on NEAT (210 $\mu$ eV full width at
half maximum (FWHM) resolution at the elastic peak). At $T=2$ K,
only the ground state is populated and therefore all excitations
arise from the ground state doublet $\left|S=4,M_S=\pm4\right>$. We
observed a strong transition at 1.77(2) meV, which we assign to the
intra-multiplet transition to the $\left|S=4\,M_S=\pm3\right>$ level,
in agreement with FDMR results (see above). One further excitation
was observed at higher energy at 2.53(1) meV.\\\indent At $T=20$ K,
we detected additional excitations at 1.05(1) meV and 1.31(1) meV,
which must be due to transitions from excited states. All peaks in
the INS spectra show a very unusual asymmetric line-shape, which we
assign to lattice solvent loss (see above).
From the comparison of INS data with the FDMR results, we can deduce
that the excitation at 2.53 meV has a pure inter-multiplet origin,
being absent in the FDMR spectra (see Fig. \ref{fig:FDMR_Mn6_S4}).
This is also confirmed by the $Q$-dependence of the scattering
intensity of the observed excitations. Figure \ref{fig:Mn6_S4} (b)
shows this dependence for the $\left|S=4 ,
M_S=\pm4\right>\rightarrow\left|S=4\,M_S=\pm3\right>$ and
$\left|S=4,M_S=\pm4\right>\rightarrow\left|S=3,M_S=\pm3\right>$
transitions. A characteristic oscillatory behavior has been observed
for the $Q$ dependence of the inter-multiplet INS transition (black
squares), which presents a maximum of intensity at a finite $Q$
value (that is related to the geometry of the molecule), and
decreasing intensity as $Q$ goes toward zero. This Q dependence is
typical for magnetic clusters and reflects the multi-spin nature of
the spin states \cite{Furrer77, Waldmann03b}. By contrast, the intra-multiplet
excitation (green circles) has maximum intensity at $Q=0$, as expected for a transition with $\Delta S =0$, 
and the intensity decreases with increasing $Q$, following the magnetic form
factor.
\\\indent The INS data directly reveal the presence of low lying
excited multiplets. Indeed, the difference in energy between the
lowest and the highest energy levels of the anisotropy split $S=4$
ground state is given, as a first approximation,  by $|D|S^2$=4.2 meV.
The presence of an inter-multiplet excitation at only 2.53 meV
energy transfer, therefore below 4.2 meV, indicates that the first
excited $S$ multiplet lies within the energy interval of the
anisotropy split $S=4$ state. This suggests that the observed low
energy excitations are possibly not pure intra-multiplet transitions, but are
expected to originate from the $S=4$ ground state and from the first
excited $S$ multiplet. Therefore the exact assignment of those
excitations requires a more accurate analysis beyond the GSH
approximation. Indeed, one fundamental requirement for the validity
of the GSH approximation, i.e. an isolated ground state well
separated from the the excited states, is not fulfilled and $S$ is
not a good quantum number to describe the ground state of the
molecule. To model the data it is thus necessary to use the full
microscopic spin Hamiltonian of Eq. \ref{eq:H_micro}. \\Given the
low symmetry of the triangular units in ({\bf 1}), the number of
free parameters in Eq. \ref{eq:H_micro} would be too large to obtain
unambiguous results, considering the low number of experimentally
observed excitations. Hence, we have chosen to describe the
low-energy physics of ({\bf 1}) by a simplified dimer model, an
approximation which has already previously been adopted for ({\bf
2}) and ({\bf 3}) (see Ref. \onlinecite{Bahr2008}). More
specifically, the two triangular units are described as two
ferromagnetically-coupled $S=2$ spins, which also experience an
effective uniaxial crystal-field potential:
\begin{eqnarray}
H_{dimer}=J ({\bf S}_A \cdot {\bf S}_B)+d
(S_{A,z}^2+S_{B,z}^2).
    \label{eq:H_dimer}
\end{eqnarray}
The spin Hamiltonian has been diagonalized numerically and the $J$ and $d$
parameters have been varied to obtain a best fit of the experimental
data. The position of the peak at 1.77 meV does not depend on the
exchange interaction, therefore its position sets the value of the
axial anisotropy $d$ parameter. Given the $d$ parameter, a fit of
the position of the peak at 2.53 meV sets the isotropic exchange
parameter $J$.
\begin{figure}[htb!]
\includegraphics[width=8cm,angle=0]{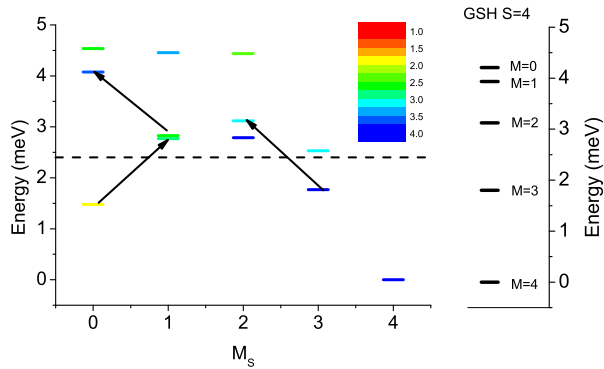}
\caption{(Color online) Calculated energy level diagram for molecule (\textbf{1}).
The level scheme on the left side is calculated using the dimer
model spin Hamiltonian in Eq. \ref{eq:H_dimer}. The color maps $S_{\text{eff}}$ , where
$\langle S^2\rangle := S_{\text{eff}}(S_{\text{eff}}+ 1)$. The black
dashed line corresponds to the observed value of $U_{\text{eff}}$ = 28 K. The black arrows
indicate transitions which contribute to the observed peak in the INS and FDMR spectra
at $E\approx 1.33$ meV (see text for details). The level diagram on the right has been calculated using the GSH
approximation (Eq. \ref{eq:GSH}).} \label{EnergyLevelS=4}
\end{figure}

The best fit of the experimental data is obtained with $J=-0.19$ meV
and $d=-0.59$ meV. The calculated energy level scheme is reported in
Fig. \ref{EnergyLevelS=4} (left), where the comparison with the
energy level diagram in the GSH approximation is also reported
(right). The value of S$_{\text{eff}}$ (where $\langle S^2\rangle :=
S_{\text{eff}}(S_{\text{eff}}+ 1)$) is labeled in color and shows
that the first $S=3$ excited state is completely nested within the
$S=4$ ground state. From Fig. \ref{EnergyLevelS=4} it is also clear
that the GSH model does not account for a number of spin states
different from the ground state $S=4$ multiplet at low energy.
Furthermore, the assignment of the observed excitations can be
misleading if considering the GSH approximation only. For example,
using the GSH model, the observed peak at 1.33 meV can only be
attributed to a pure intra-multiplet excitation from
$\left|4,\pm3\right>$ to $\left|4,\pm2\right>$, whilst using Eq.
\ref{eq:H_dimer}, it is found to be a superposition of several
inter-multiplet and intra-multiplet transitions (indicated by arrows
in Fig. \ref{EnergyLevelS=4}). The GSH approximation fails to
describe the low energy level diagram of the molecule and
consequently fails to describe the relaxation of the magnetization.
Indeed, the presence of excited states nested within the ground
state multiplet has a significant effect on the relaxation dynamics,
as discussed in section \ref{Discussion}.

\subsection{Mn$_6$ (\textbf{2}) $U_{\text{eff}}\approx$ 53 K vs. Mn$_6$ (\textbf{3}) $U_{\text{eff}}\approx$ 86.4 K}

\begin{figure}[htb!]
\includegraphics[width=8.5cm,angle=0]{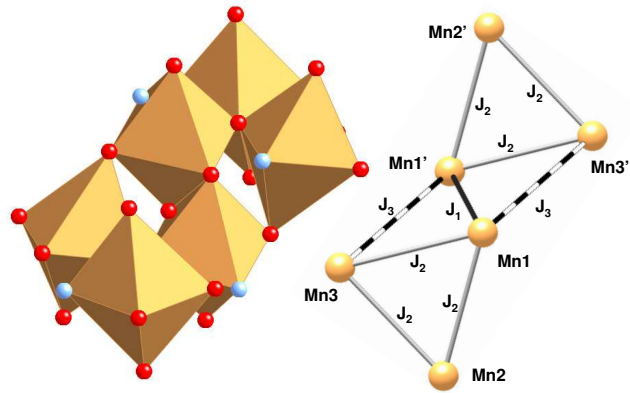}
\caption{(Color online) Structure of the Mn$_{6}$ (\textbf{2})
molecular core. The Mn$^{3+}$ ions are located at the vertices of
two oxo-centered triangles. All Mn ions are in octahedral geometry
and the octahedra are highlighted in orange (left figure). Color
scheme: Mn, orange, O, red,  N, light blue. H and C ions are omitted
for clarity. On the right, a schematic representation is given, together with the exchange coupling scheme
adopted for the spin Hamiltonian calculations.}
\label{fig:Mn6_S12_poly}
\end{figure}

\begin{figure}[htb!]
\includegraphics[width=8cm]{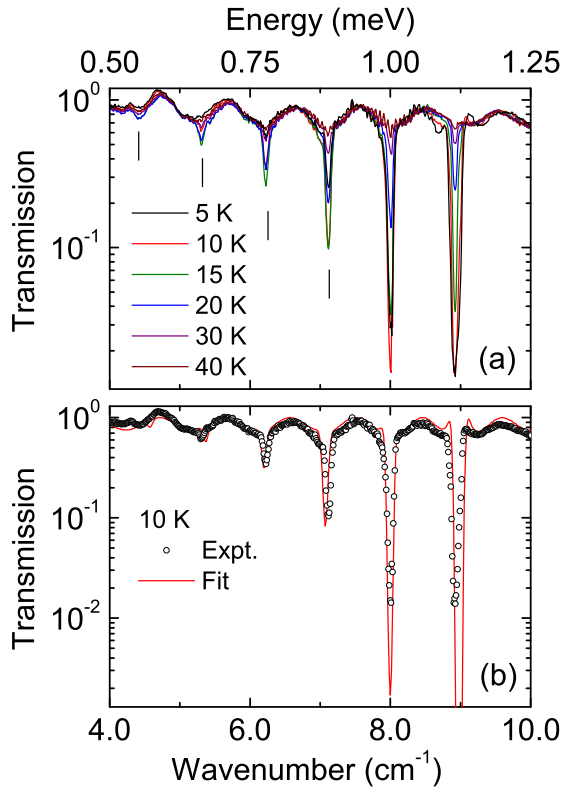}
\caption{(Color online) (a) FDMR spectra of unpressed
polycrystalline powder of (\textbf{3}) recorded at various
temperatures. (b) Experimental and fitted spectrum at $T = 10$
K.} \label{FDMR86K}
\end{figure}

\begin{figure}[htb!]
\includegraphics[width=6cm]{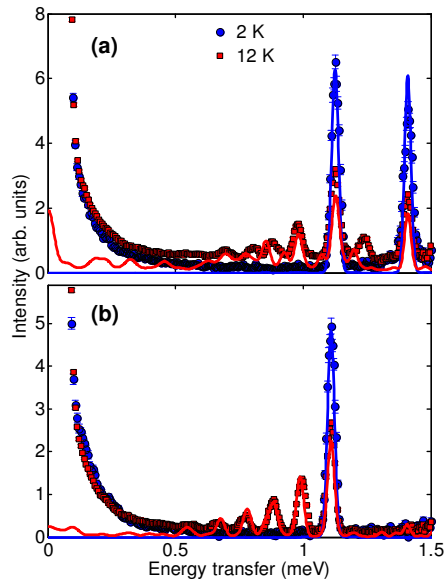}
\caption{(Color online) INS spectra collected on IN5 with incident
wavelength of 6.7 {\AA} at $T=2$ K (blue circles) and $T=12$ K (red
squares). (a) for sample (\textbf{2}) and (b) for sample
(\textbf{3}). The spectra calculated with the parameters listed
in Table \ref{table:couplings} are shown as continuous
lines.}\label{fig:Mn6_S12_L67_both}
\end{figure}

\begin{figure}[htb!]
\includegraphics[width=6cm]{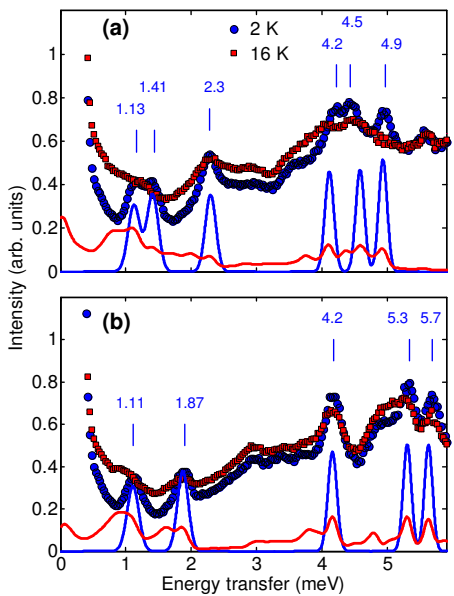}
\caption{(Color online) INS spectra collected on IN5 with incident
wavelength of 3.4 {\AA} at $T=2$ K (blue circles)  and 16 K (red squares). (a) for
sample (\textbf{2}) and (b) for sample (\textbf{3})
. The observed transitions are labeled with the corresponding transition energies in meV.}\label{fig:Mn6_S12_L34_both}
\end{figure}

Introducing sterically more demanding oximate ligands results in a
twisting of the Mn-N-O-Mn torsion angle \cite{Milios07Ueff53}, which
causes switching of the intra-triangle exchange interactions from
antiferromagnetic to ferromagnetic, resulting in a large increase of
the spin of the ground state from $S=4$ to $S=12$. Here, we study two
((\textbf{2}) and (\textbf{3}), respectively) of the many published
derivatives of these $S=12$ Mn$_6$ clusters. Compound (\textbf{2}) has
undergone two structural changes compared to (\textbf{1}). First of
all, the distance between the phenolato oxygen and the two square
pyramidal Mn$^{3+}$ ions has decreased from $\approx 3.5$ \AA\ to
$\approx 2.5$ \AA, thus all Mn$^{3+}$ ions are now in
six-coordinated distorted octahedral geometry (see Fig.
\ref{fig:Mn6_S12_poly}). Secondly, the torsion angles of the
Mn$-$N$-$O$-$Mn moieties has increased strongly with respect to
those in (\textbf{1}), being 38.20$^\circ$, 39.9$^\circ$ and
31.26$^\circ$, compared to 10.7$^{\circ}$, 16.48$^{\circ}$ and 22.8$^{\circ}$ for
(\textbf{1}). In (\textbf{3}), the introduction of two methyl groups on
the carboxylate ligand has increased the non-planarity of the
Mn$-$N$-$O$-$Mn moieties further, giving torsion angles of
39.08$^\circ$, 43.07$^\circ$ and 34.88$^\circ$
\cite{Milios07Ueff86}. The result is that the weakest ferromagnetic
coupling is significantly stronger for (\textbf{3}) compared to (\textbf{2}).
Using a single $J$ model (e.g. assuming that the intra- and
inter-triangle exchange couplings are equal), Milios \textit{et al.}
fitted the DC susceptibility data for molecules
(\textbf{2}) and (\textbf{3}) and obtained: $J$(\textbf{2})$=-0.230$ meV and $J$(\textbf{3})$=-0.404$ meV, respectively \cite{Milios07magstruc,
InglisDalton09} (in our notation for the spin Hamiltonian).\\
In spite of the fact that both (\textbf{2}) and (\textbf{3}) have $S=12$ ground states and similar geometrical structures, radically different effective energy barriers towards the relaxation of the magnetization were observed, being $U_{\text{eff}}\approx 53$ K for (\textbf{2}) and $U_{\text{eff}}\approx 86.4$ K for (\textbf{3}). Here, we aim to understand this difference by an in-depth study of the energy level structure by means of FDMR and INS.\\
\begin{figure}[htb!]
\includegraphics[width=8cm]{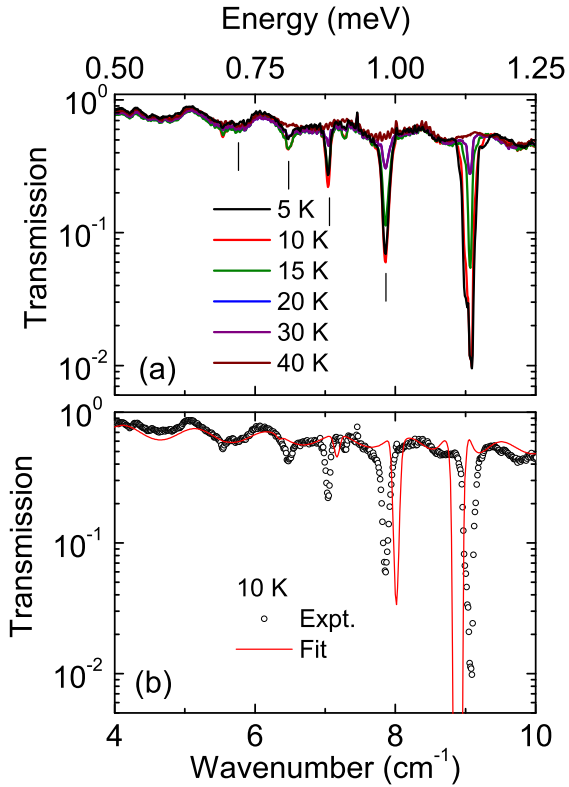}
\caption{(Color online) (a) FDMR spectra recorded on a pressed
powder pellet of 2 at different temperatures. At the lowest
temperature, the highest-frequency line has highest intensity. The
other transitions are indicated by vertical lines. (b) 10K FDMR
spectrum (symbols) and best fit using the GSH (Eq. \ref{eq:GSH}), with $D =
-0.368$ cm$^{-1}$, and $B_4^0 = -4.0 \times 10^{-6}$ cm$^{-1}$.}
\label{FDMR53K}
\end{figure}
Figure \ref{FDMR53K} shows FDMR spectra recorded on a pressed
powder pellet of (\textbf{2}) at different temperatures. The baseline
shows a pronounced oscillation, which is due to Fabry-P\'erot-like
interference within the plane-parallel pellet \cite{kirchner07}. The
oscillation period and downward slope to higher frequencies are
determined by the thickness of the pellet and the complex dielectric
permittivity, which were determined to be $\epsilon'=3.01$ and
$\epsilon''=0.049$, values typical for molecular magnet samples. In
addition, five resonance lines are observed which we attribute to
resonance transitions within the $S=12$ multiplet. Thus, the highest
frequency line is assigned to the
$\left|12,\pm12\right>\rightarrow\left|12,\pm11\right>$
transition, and so on. The lines are much narrower (11 $\mu$eV FWHM)
than those observed for other SMMs, e.g. 23 $\mu$eV FWHM for
Mn$_{12}$Ac. The fit procedure showed that the lines are
inhomogeneously broadened and best described by Gaussian lineshapes. The small
linewidth indicates that distributions in ZFS parameters ($D$-strain) are
small in these samples. A fit of the GSH parameters (Eq. \ref{eq:GSH}) to the
observed resonance frequencies, yields $D_{S=12}=-0.368$ cm$^{-1}$
(0.0456 meV) and $B_4^0=-4.0\times10^{-6}$ cm$^{-1}$
(4.96$\times10^{-7}$ meV) best parameter values. The theoretical
energy barrier calculated from these ZFS parameters is
$U_{\text{theor}}=76$ K, which is much larger than the
experimentally found $U_{\text{eff}}\approx$ 53 K, indicating that
the molecule can shortcut the barrier in some way. The ZFS values
are in themselves not remarkable, and close to those reported for
other manganese clusters with similar ground state spins, e.g.
$D_{S=10} = -0.457$ cm$^{-1}$ for Mn$_{12}$Ac \cite{Mirebeau99},
$D_{S=\frac{17}{2}} = -0.247$ cm$^{-1}$ for Mn$_{9}$
\cite{Piligkos05}. Interestingly, the fourth order axial ZFS is an
order of magnitude smaller than for Mn$_{12}$Ac. This type of ZFS is
currently accepted to parametrize effects of mixing between spin
multiplets ($S$-mixing) \cite{CarrettaPRL04}, which would mean
that $S$-mixing is only limited, contrary to expectation.
However, the fit does not simulate the resonance line positions and
intensities satisfactorily, which is in contrast to the situation
for other molecular nanomagnets that feature strong $S$-mixing, e.g.
Ni$_4$ \cite{Kirchner08, Sieber05}. Therefore, the investigated Mn$_6$ SMM represents an
example where the giant spin model cannot satisfactorily describe
FDMR spectra, and it will be shown below that this is due to a
complete breakdown of the giant spin model. It will also be shown
that the resonance line at 0.80 meV is due to a transition within
the $S = 11$ excited multiplet. However, removal of this resonance
line does not result in a better fit. The calculated line
intensities are much larger than those experimentally found,
especially for the highest-frequency lines. This we attribute to a
combination of parasitic radiation in the cryostat, and the presence
of many more states than taken into account by the giant spin model,
which decreases the relative population for any given state.
\\\indent
Similar FDMR results were obtained for (\textbf{3}) (Fig.
\ref{FDMR86K}), and six sharp resonance lines were observed.
A fit of the GSH parameters to the observed resonance line positions
yields the following values: $D = -0.362\pm0.001$ cm$^{-1}$ (-0.0449
meV), and $B_4^0 = -6.0\pm0.4 \times 10^{-6}$  cm$^{-1}$
(-7.4$\times 10^{-7}$ meV). The simulated spectrum matches the
experiment much more closely for (\textbf{3}), especially for the
high-frequency lines. Interestingly, the theoretical energy barrier
($U_{\text{theor}}=75$ K) is virtually the same as for (\textbf{2}),
but \emph{smaller} than the experimentally found energy barrier
($U_{\text{eff}}=86$ K). This unprecedented finding means that the
magnetization relaxation must involve states that do not belong to
the ground spin multiplet \cite{Carretta08}. Again, we turn to INS
to determine the positions of the excited spin multiplets, which
will allow full characterization of the system.
\begin{table}
\squeezetable \caption{INS and FDMR peak positions of the observed
excitations for  (\textbf{2}) and (\textbf{3}) (in meV).}
\label{table:list}
\begin{ruledtabular}
\begin{tabular}{lll|lll}
  (\textbf{2}) & INS         & FDMR  &  (\textbf{3})        &   INS        & FDMR   \\
\hline
&   4.9(2)  & n.o.\footnote{not observed} &  & 5.7(2)   & n.o.        \\
&   4.5(1)  & n.o.                        &  & 5.3(2)   & n.o.        \\
&   4.2(2)  & n.o.                        &  & 4.2(2)   & n.o.        \\
&   2.3(2)  & n.o.                        &  & 1.87(3)  & n.o.        \\
&  1.41(2)  & n.o.                        &  & 1.11(1)  & 1.107(7)    \\
&  1.24(7)  & n.o.                        &  & 0.99(1)  & 0.993(6)    \\
&  1.13(2)  & 1.127(5)                    &  & 0.88(2)  & 0.883(6)    \\
&  0.98(2)  & 0.975(5)                    &  & 0.77(1)  & 0.772(7)    \\
&  0.88(3)  & 0.873(6)                    &  & 0.66(1)  & 0.657(7)    \\
&  0.80(2)  & 0.803(7)                    &  & 0.55(2)  & 0.551(10)   \\
&  0.70(2)  & 0.687(5)                    &  & 0.48(1)  & n.o.        \\
&  0.57(4)  & n.o.                        &  & 0.45(1)  & n.o.        \\
&           &                             &  & 0.34(1)  & n.o.        \\
&           &                             &  & 0.31(1)  & n.o.        \\
&           &                             &  & 0.25(1)  & n.o.        \\
&           &                             &  & 0.21(3)  & n.o.        \\
\end{tabular}
\end{ruledtabular}
\end{table}
Figures (\ref{fig:Mn6_S12_L67_both}a) and (\ref{fig:Mn6_S12_L67_both}b) show
the high resolution INS experimental data for compounds (\textbf{2})
and (\textbf{3}), respectively, collected on IN5 with an incident
wavelength of 6.7 {\AA} (53 $\mu$eV FWHM resolution at the elastic peak).
At the lowest temperature $T = 2$ K only the ground state is
populated and, due to the INS selection rules, only transitions with
$\Delta S = 0,\pm 1$ and $\Delta M = 0,\pm 1$ can be detected. The
lowest energy excitation can be thus easily attributed to the
intra-multiplet transition from the $\vert S=12,M_S=\pm12\rangle$
ground state to the $\vert S=12,M_S=\pm 11\rangle$ first excited
level. The position of this intra-multiplet excitation is found to
be at about the same energy in both compounds, i.e. $\sim 1.1$ meV,
indicating only small differences in the anisotropy of the two systems. In contrast the first
inter-multiplet $S=12\rightarrow S=11$ excitation at about 1.41 meV
in compound (\textbf{2}) is not visible in the spectra at 6.7 {\AA}
of compound (\textbf{3}). This can be understood looking at the data
at higher energy transfer, collected with an incident wavelength of
3.4 {\AA} (see Figs. (\ref{fig:Mn6_S12_L34_both}a) and
(\ref{fig:Mn6_S12_L34_both}b)). Indeed the first inter-multiplet
excitation is considerably raised in energy in compound (\textbf{3})
with respect to compound (\textbf{2}), from 1.41 meV to 1.87 meV.
This gives a direct evidence of an increase of the isotropic
exchange parameters, while the anisotropic parameters are
approximately the same for both molecules. The INS spectra collected
at a base temperature of 2 K, enabled us to directly access the
whole set of intra-multiplet and inter-multiplet transitions allowed
by the INS selection rules in both compounds. By raising the
temperature to 16 K the intensity of the magnetic peaks decreases,
thus confirming their magnetic origin. A total of five
inter-multiplet excitations for compound (\textbf{2}) toward
different $S=11$ excited states can be detected. For compound
(\textbf{3}) four inter-multiplet excitations have been observed.
All the magnetic excitations are marked in Fig.
\ref{fig:Mn6_S12_L34_both} with the corresponding transition
energies.

\begin{figure}[htb!]
\includegraphics[width=8cm]{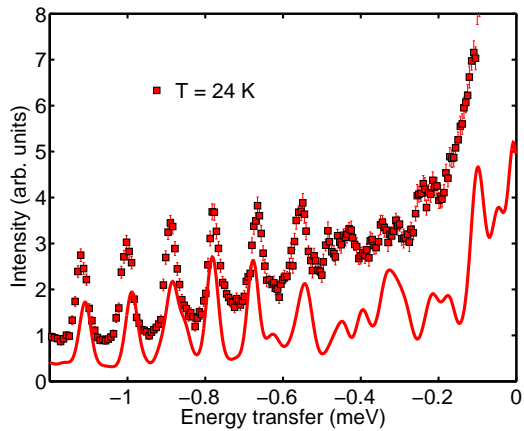}
\caption{(Color online) High resolution INS spectra of molecule
(\textbf{3}) collected on IN5 with incident wavelength 10.5 {\AA}
 at 24 K. The
energy gain spectra is displayed. Continuous lines are the
calculations using the spin Hamiltonian of Eq. (\ref{eq:H_micro}).
}\label{fig:Mn6_S12_U86_85A_105A}
\end{figure}

To complete our investigations of the transitions within the $S=12$
ground-state multiplet, we additionally performed high resolution
measurements of molecule (\textbf{3}) using IN5 with incident
wavelengths of 10.5 \AA\ (FWHM = 13
$\mu$eV at the elastic line)(see Fig.
\ref{fig:Mn6_S12_U86_85A_105A}). These measurements allowed us to
observe transitions originating from the top of the anisotropy
barrier.

\begin{figure}[htb!]
\includegraphics[width=6cm]{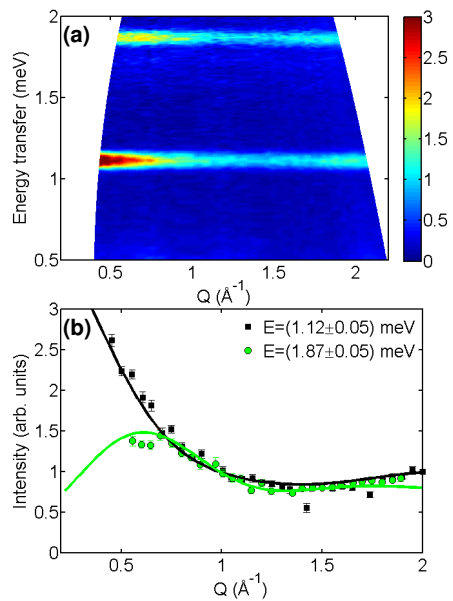}
\caption{(Color online) (a) Energy-wavevector colormap of sample
(\textbf{3}) collected on IN5 with incident wavelength of 5.0 \AA.
(b) $Q$-dependence of two transitions from the ground state. The
black squares correspond to the $|S=12,M_s=\pm12> \rightarrow
|S=12,M_s=\pm11>$ intra-multiplet transition and the green circles
display the $|S=12,M_s=\pm12 \rightarrow
|S=11,M_s=\pm11>$.}\label{fig:Mn6_U86_L5_qdep}
\end{figure}

\begin{figure}[htb!]
\includegraphics[width=7cm,angle=0]{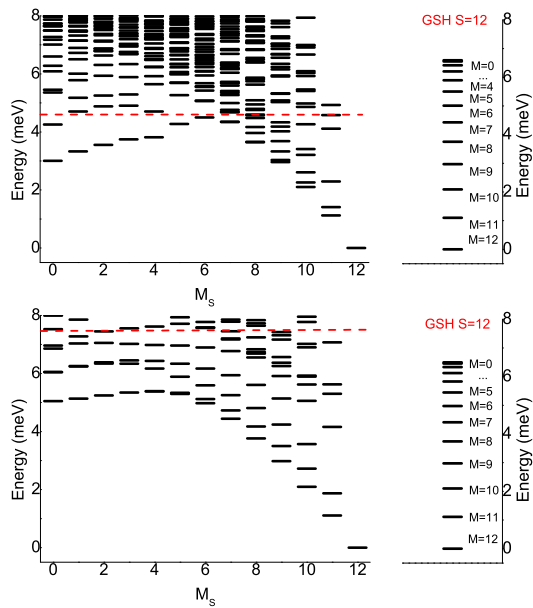}
\caption{(Color online) Calculated energy level diagram for molecule (\textbf{2})
(top) and molecule (\textbf{3}) (bottom). The level scheme on the
left side is calculated using the microscopic spin Hamiltonian in Eq. \ref{eq:H_micro},
while the level diagram on the right has been calculated in
the GSH approximation (Eq. \ref{eq:GSH}). The dashed lines correspond to the observed value of $U_{\text{eff}}$.}
\label{EnergyLevel}
\end{figure}

A further confirmation of the good assignment of the observed
excitations is provided by the study of their $Q$-dependence. As
revealed by Fig. \ref{fig:Mn6_U86_L5_qdep}, the intra-multiplet
transition ($\Delta S$=0) shows a distinctive $Q$-dependence, with a
pronounced intensity at low $Q$, that dies out quite rapidly following
the Mn$^{3+}$ form factor. In contrast, inter-multiplet excitations
present flatter behavior, with considerably less intensity at
low $Q$.
The assignment of the observed excitations to intra-multiplet or
inter-multiplet transitions has been confirmed by comparison with
FDMR measurements performed on both compounds (see Fig.
\ref{FDMR53K} and Fig. \ref{FDMR86K}). The position of the
intra-multiplet INS transitions are consistent with the FDMR
measurements performed on the same sample (see Table
\ref{table:list}). Due to the different selection rules of INS and
FDMR, we can conclude that all the peaks observed at T = 2 K above
1.2 meV energy transfer correspond to inter-multiplet transitions,
since they are absent in the FDMR spectra.\\The straightforward
assignment of the base temperature observed excitations allows us to
draw some considerations on the experimentally deduced energy level
diagram. For both compounds, a rough estimate of the splitting of
the spin ground multiplet gives $|D|S^2 \simeq 6.5$ meV. This value is
comparable to the energy interval explored by the high energy
transfer INS data (Fig. \ref{fig:Mn6_S12_L34_both}), where most of
the inter-multiplet $S=12 \rightarrow S=11$ excitations have been
observed. This experimental observation leads to the conclusion that
also in (\textbf{2}) and (\textbf{3}) several excited states lie
within the anisotropy split ground state, with the consequent
breakdown of the GSH approximation. Due to the inadequacy of the GSH
for (\textbf{2}) and (\textbf{3}), the microscopic spin Hamiltonian
(Eq. \ref{eq:H_micro}) was used to model the data and extract the
exchange constants and anisotropies . The minimal set of free
parameters is given by three different exchange constants
$J_{11^\prime}\equiv J_1$, $J_{12}=J_{23}=J_{13}=J_{1^\prime
2^\prime}=J_{2^\prime 3^\prime}=J_{1^\prime 3^\prime}\equiv J_2$,
and $J_{13^\prime}=J_{1^\prime 3}\equiv J_3$ (Fig.
\ref{fig:Mn6_S12_poly}) and two sets of crystal-field (CF)
parameters $d_1=d_{1^\prime}$, $c_1=c_{1^\prime}$, and
$d_2=d_{2^\prime}$, $c_2=c_{2^\prime}$. Indeed, the ligand cages of
sites 1 and 3 are rather similar and we assumed the corresponding CF
parameters to be equal. Since experimental information is
insufficient to fix independently the two small $c$ parameters,we
have chosen to constrain the ratio
$c_1/c_2$ to the ratio $d_1/d_2$.\\
The isotropic exchange and crystal field parameters deduced by the
simultaneous best fit of the experimental data are reported in Table
\ref{table:couplings}. Figure \ref{EnergyLevel} shows the calculated energy level diagram using the best fit procedure for Eq. \ref{eq:H_micro} (left) and the GSH model (right) for  (\textbf{2}) and (\textbf{3}).

\begin{table}
\caption{Isotropic exchange and CF parameters for Eq. \ref{eq:H_micro} (in meV)
deduced by fitting INS and FDMR data for the two Mn$_6$ S=12
compounds.} \label{table:couplings}
\begin{ruledtabular}
\begin{tabular}{rcrrrrrrrr}
 &    U$_{eff} (K)$ & $J_1$ & $J_2$ & $J_3$ & $d_1$ & $d_2$ & $c_1$ \\
\hline
(\textbf{2}) & 53 &-0.61  & -0.31  & 0.07   & -0.23   & -0.97   & -0.0008    \\
(\textbf{3}) & 86.4 & -0.84  & -0.59  & 0.01   & -0.20   & -0.76   & -0.001 \\
\end{tabular}
\end{ruledtabular}
\end{table}

\section{Discussion}\label{Discussion}

The experimental data collected on the three variants of Mn$_6$
clusters provide direct evidence that a general feature for this
class of compounds is the nesting of excited multiplets within the
ground state multiplet. This is an unavoidable effect when the
isotropic exchange parameters have the same order of magnitude as
the single ion anisotropy parameters, as it happens to be for
Mn$_6$. The nesting of spin states can be clarified by observing the
energy level diagrams for the three molecules presented in Fig.
\ref{EnergyLevelS=4} and Fig. \ref{EnergyLevel}. The diagram on the
left shows the energy levels calculated by a diagonalization of the
full spin Hamiltonian, while the energy level scheme on the right
hand side has been calculated considering the GSH approximation.  It
is clear that the GSH does not account for any of the spin states
with $S$ different from $S_{\text{GS}}$ that lie within the split GS
energy level diagram. The above states represent a shortcut for the 
relaxation of the magnetization and can promote resonant inter-multiplet
tunneling processes that manifest as additional steps in the magnetization
curve absent in the GS model\cite{Ramsey08, Bahr2008, Yang07, Soler03, Carretta09poly}.
The overall result is a lowering of the
effective anisotropy barrier with respect to an ideal molecule where
the spin ground state is well separated from the excited ones, 
as was firstly demonstrated in Ref. \onlinecite{Carretta08}.\\
We have calculated the relaxation dynamics of molecule (\textbf{1})
following the same procedure adopted in Ref. \onlinecite{Carretta08}
for molecules (\textbf{2}) and  (\textbf{3}). We applied a master
equations formalism in which the magnetoelastic (ME) coupling is
modeled as in Ref. \onlinecite{CarrettaPRL06}, with the quadrupole
moments associated to each triangular unit isotropically coupled to
Debye acoustic phonons.

The transition rates $W_{st}$ between pairs of eigenlevels of the
dimer spin Hamiltonian Eq. \ref{eq:H_dimer} is given by:
\begin{equation}
W_{st} =
\gamma^2\Delta_{st}^3n(\Delta_{st})\!\!\!\!\sum_{A,B,q_1,q_2}\!\!\!\!\langle
s|O_{q_1,q_2}({\bf S}_A)|t\rangle \overline{\langle
s|O_{q_1,q_2}({\bf S}_B)|t\rangle}
\end{equation}
where $O_{q_1,q_2}({\bf S}_{A,B})$ are the components of the
Cartesian quadrupole tensor operator, $n(\Delta_{st}) = (e^{\hbar
\Delta_{st}/k_BT}-1)^{-1}$ and $\Delta_{st}=(E_s-E_t)/\hbar$. We
found out that the resulting relaxation spectrum at low $T$ is
characterized by a single dominating relaxation time whose
$T$-dependence displays a nearly Arrhenius behavior  $\tau = \tau_0
\exp(U/k_BT)$, as previously observed for molecules (\textbf{2}) and
(\textbf{3}) \cite{Carretta08}. The relaxation dynamics of $M$ is
indeed characterized by two separated time scales: fast processes
that determine the equilibrium within each well of the double-well
potential and a slow inter-well process that at low temperature
determines the unbalancing of the populations of the two wells, and
thus sets the time scale for the reversal of the magnetization. As
can be observed from the energy level diagram of Fig.
\ref{EnergyLevelS=4} there are several levels that can be involved
in the inter-well relaxation process, giving rise to an overall
effective barrier $U_{\text{eff}}$ different from the simple energy
difference between the $M=0$ and $M=\pm 4$ states. The corresponding
calculated energy barrier $U_{\text{calc}}$= 32 K reproduces quite
well the experimental value, $U_{\text{eff}}$= 28 K. The lowering of
the barrier is therefore attributed to the presence of these extra
paths. Indeed, the calculations for artificially isolated $S = 4$
yield $U= 47$ K.\\ 
It is worth commenting also on the $D$ value for
the ground state of each molecule. Whilst no large difference
between the local $d$ of the low (\textbf{1}) and high ((\textbf{2})
and (\textbf{3})) spin molecules is expected, the overall $D$ value,
as determined using the GSH approximation,  is much higher for the
$S=4$ molecule ($D\approx-0.263$ meV) than for the high spin
molecules ($D\approx-0.045$ meV ). However, this observation should
not be misinterpreted. The difference arises from the fact that $D$
 depends on the projection of the individual single-ion anisotropies of each magnetic ion onto the total spin quantum number
 $S$. In the case where the $S$-mixing is negligible and the spin ground state is a good quantum number, the $D$ parameter for a specific state $S$ can
 be written as linear combination of the single-ions anisotropy tensors (Ref. \onlinecite{Bencini90}):
\begin{equation}
{\bf D}=\sum_{i=1}^N  a_i {\bf d}_i
    \label{eq:D_sum}
\end{equation}
 The projection coefficients $a_i$ of the
 single ion anisotropy to spin states of different $S$ values can differ
 significantly, giving rise to considerably different $D$ values.
 The ligand field study of various members of the Mn$_6$
family (Ref. \onlinecite{Piligkos08}) provides experimental evidence of
this. Recent theoretical
 studies proposed that the intrinsic relationship between $S$ and $D$ causes a scaling of $U$ that goes approximately with $S^0$
 (see Ref. \onlinecite{Waldmann07} and \onlinecite{Ruiz08}), raising the question whether it is worth trying to increase
 the value of spin ground state to obtain a larger energy barrier. Indeed,
 higher spin ground states would correspond to lower $D$
 parameters, neutralizing the overall effect on the height of the
 anisotropy barrier. In recently performed electron paramagnetic resonance studies the authors proposed that the barrier goes roughly with $S^1$ instead \cite{Datta09}. In the specific case of Mn$_6$, because of the very large $S$-mixing, the
 projection onto a well defined spin state is no more justified and it is not possible to associate the barrier $U$ to a defined $S$ value.
However, if we consider the effective anisotropy barrier
 for artificially isolated $S\!=\!4$ and $S\!=\!12$ states (i.e. $U=47$ K for (\textbf{1}) and $U = 105$ K for (\textbf{2})),
we can confirm that the barrier does not go quadratically with $S$,
as one could naively deduce from the equation $U=|D|S^2$. Indeed,
U$_{S=12}$/U$_{S=4}$ = 2.2 $\ll 12^2/4^2$=9. This confirms what has
been pointed out in Ref. \onlinecite{Waldmann07}, i.e. even though
the highest anisotropy barrier is obtained with the molecule with
the highest spin ground state, the increase of the total spin is not
as efficient as one would expect and alternative routes, like
increasing the single ion anisotropy, should be considered.

\section{Conclusion}

We have performed INS and FDMR measurements on three variants of
Mn$_6$ molecular nanomagnets, which have the same magnetic core and
differ by slight changes in the organic ligands. INS measurements
have unambiguously evidenced the presence of low lying excited
states in all the three molecules. The combination of the two
techniques enabled us to determine the spin Hamiltonian
parameters used for the analysis of the magnetic properties. The
nesting of excited states within the ground state multiplet strongly
influences the relaxation behavior and plays a crucial role in
lowering the effective energy barrier. The calculations of the
relaxation dynamics give results that are consistent with the
experimental values and show that the highest barrier is obtained
for ideal molecules with an isolated ground state. This observation
might be valid for a wider class of SMMs.

\begin{acknowledgments}
This work was partly supported by EU-RTN QUEMOLNA Contract No.
MRTN-CT-2003-504880, the German Science Foundation DFG, and EPSRC. This work utilized facilities
supported in part by the National Science Foundation under Agreement
No. DMR-0454672.
\end{acknowledgments}

\end{document}